\documentclass[aps,preprint,nofootinbib,superscriptaddress,prc]{revtex4}
\usepackage{amssymb}
\usepackage{CJK}
\usepackage{indentfirst}
\usepackage{amsmath}
\usepackage{xcolor}

\usepackage{epsfig}
\usepackage[bookmarksnumbered,bookmarksopen,colorlinks,citecolor=blue,linkcolor=blue] {hyperref}
\begin{document}

\date{\today}

\title{Unified Royer Law Revision for $\alpha$-Decay Half-Lives: Shell Corrections, Pairing, and Orbital-Angular-Momentum}

\author{Kai Ren}
\affiliation{Department of Physics, Guangxi Normal University, Guilin 541004, People's Republic of China }

\author{Pengfei Ma}
\affiliation{Department of Physics, Guangxi Normal University, Guilin 541004, People's Republic of China }

\author{Minghui Hu}
\affiliation{Department of Physics, Guangxi Normal University, Guilin 541004, People's Republic of China }

\author{Junlong Tian}
 \email{tianjl@gxnu.edu.cn}
 \affiliation{Department of Physics, Guangxi Normal University, Guilin 541004, People's Republic of China }
\affiliation{Guangxi Key
Laboratory of Nuclear Physics and Technology,  Guilin 541004, People's Republic of China}

\begin{abstract}

The Royer law is a widely used empirical relation for calculating $\alpha$-decay half-lives but requires 12 parity-dependent parameters. It exhibits systematic deviations near the $N = 126$ shell closure. We propose an improved Royer law by adding a shell-correction term, an odd-even pairing indicator, and an orbital-angular-momentum contribution. This unified framework reduces the number of free parameters to just four, leading to significant improvements in accuracy. The root-mean-square deviation across 550 experimental data points decreases from 0.520 to 0.279, corresponding to a 66.7$\%$ reduction in parameters and a 46.3$\%$ improvement in accuracy. Using this refined formalism, we predict $\alpha$-decay half-lives for superheavy nuclei with atomic numbers $Z = 117-120$.

\end{abstract}

\pacs{21.10.Dr, 23.40.-s, 21.65.Ef}
 %24.10.-i Nuclear reaction models and methods
 %21.65.Ef Symmetry energy
 %24.70.+s Polarization phenomena in reactions
 %25.45.-z 2H-induced reactions
 % PACS, the Physics and Astronomy Classification Scheme.
 \keywords{mass parabola, double $\beta$-decay energies, orbital-angular-momentum}%

\maketitle

\section{\label{sec:level1}Introduction}

The prediction of $\alpha$-decay half-lives is fundamental to nuclear structure studies\cite{Seif1,2014pxa,2014yga,2015mnw,2011zz,2002zk,2015aha,1972cmm,2006bi,1987rkx,1990qu,2014zha,2003zz,2006ay,2007ev}, particularly for superheavy nuclei\cite{2007zza,2010zze,2010zzo,2003hm,2005jx,2012ju}. In 1911, Geiger and Nuttall observed that plotting  $\log_{10}T_{1/2}$ against $Q^{-1/2}$ yields a linear relationship for even-even isotopes \cite{Geiger11}. The phenomenon of $\alpha$ decay was first explained as a quantum-tunneling process by Gamow \cite{Gamow1} and independently by Gurney and Condon \cite{Gurney1}. Since then, a variety of theoretical models have been developed to deepen our understanding of $\alpha$ decay. Notable examples include the Viola–Seaborg–Sobiczewski (VSS) formula \cite{VIOLA1966741}, the effective liquid-drop model \cite{1998pz,1993iga,2022skm,1996zzd,2002nyh}, the generalized liquid-drop model (GLDM) \cite{2006dj1,Dongjm2010}, the fission-like model \cite{1993sku}, and several others \cite{2007rg,2006dj2,2003ai,2016bbw,2016jnb,2012zza}. In parallel, many empirical formulas have been proposed based on the Geiger–Nuttall law (GNL) or quantum-tunneling arguments, such as the universal decay law (UDL) \cite{2009zzb,2009id}, the Royer law \cite{2000Alpha}, the Deng–Zhang–Royer (DZR) formula \cite{2020gyy}, and the new Geiger–Nuttall law (NGNL) \cite{Ren:2012zza}. Recent studies have further advanced the systematic understanding of $\alpha$ decay. For instance, El Batoul \textit{et al.} \cite{Review1} refined empirical formulations by incorporating a position-dependent mass formalism to improve accuracy; You \textit{et al.} \cite{Review2,Review3} applied machine-learning techniques along with deformation effects to enhance predictive reliability; and Ismail \textit{et al.} \cite{Review4} investigated structural dependencies within Royer-type models.

The Royer law is a widely used empirical relation for calculating $\alpha$-decay half-lives, yet it relies on 12 parity-dependent parameters. Despite its broad application, this model exhibits significant deviations from experimental data in the region of the $N = 126$ shell closure. To address these limitations, we propose an improved Royer law that incorporates the shell-correction energy, a pairing term, and an angular-momentum term. This modification not only reduces discrepancies near \(N=126\) but also enhances the overall accuracy of calculating \(\alpha\)-decay half-lives. Our analysis is based on 550 measured $\alpha$-decay half-lives, comprising the 539 entries from the NUBASE2020 database \cite{Nubase2020} and 11 additional nuclei from recent publications ($^{170}\mathrm{Hg}$ \cite{170Hg},  $^{214}\mathrm{U}$, $^{216}\mathrm{U}$, $^{218}\mathrm{U}$ \cite{214216218U}, $^{160}\mathrm{Os}$ \cite{160Os}, $^{190}\mathrm{At}$ \cite{190At}, $^{207}\mathrm{Th}$ \cite{207Th}, $^{272}\mathrm{Hs}$, $^{276}\mathrm{Ds}$ \cite{272Hs276Ds}, $^{210}\mathrm{Pa}$ \cite{210Pa} and $^{286}\mathrm{Mc}$ \cite{286Mc}). The parameters of the improved Royer formula are determined by fitting to this robust and comprehensive dataset. The study focuses exclusively on ground-state-to-ground-state $\alpha$ decays. To ensure accurate extraction of half-lives, we account for the experimental branching ratio $R$ of $\alpha$ decay from the parent ground state to various daughter states. From NUBASE2020 \cite{Nubase2020}, we initially considered 701 nuclei with reported \(\alpha\)-decay branching ratios. After applying rigorous selection criteria---experimental uncertainties below 50\%, branching-ratio uncertainties smaller than $R$ itself, and exclusion of \(^{264}\mathrm{Hs}\) (due to \(\log_{10}T_{1/2} = 0\))—we retained 539 nuclei. These were further categorized into four parity groups: 190 even-even (e-e), 146 even-odd (e-o), 114 odd-even (o-e), and 100 odd-odd (o-o). Alternatively, the dataset can be classified by the orbital-angular-momentum $l$ of the transition. Favored $\alpha$ decays (\(l = 0\)) constitute the majority, with 406 cases (74\% of the total), distributed as 190 e-e, 88 e-o, 71 o-e, and 57 o-o. Unfavored decays (\(l \neq 0\)) account for the remaining 144 cases (26\%), comprising 58 e-o, 43 o-e, and 43 o-o.

\section{\label{sec:level2}Theoretical framework}
The Royer law \cite{2000Alpha} establishes a benchmark relationship between $\log_{10} T_{1/2}$ and nuclear properties, expressed as
\begin{equation} \label{Eq1}
	\log_{10}T_{1/2} = a + bA^{1/6}\sqrt{Z} + c\frac{Z}{\sqrt{Q}},
\end{equation}
where $A$, $Z$, and $Q$ represent the mass number, proton number, and decay energy of the parent nucleus, respectively. The parameters $a$, $b$, and $c$ are determined by fitting the experimental data. The original Royer law employs parity-dependent parameters (Table~\ref{tab:params1}). It requires separate treatments for e-e, e-o, o-e, and o-o nuclei, based on the proton ($Z$) and neutron ($N$) parity of the parent nucleus. A total of 12 adjustable parameters are distributed across these four parity groups.

\begin{figure*}[htb]
	\centering
	\includegraphics[width=0.6\linewidth]{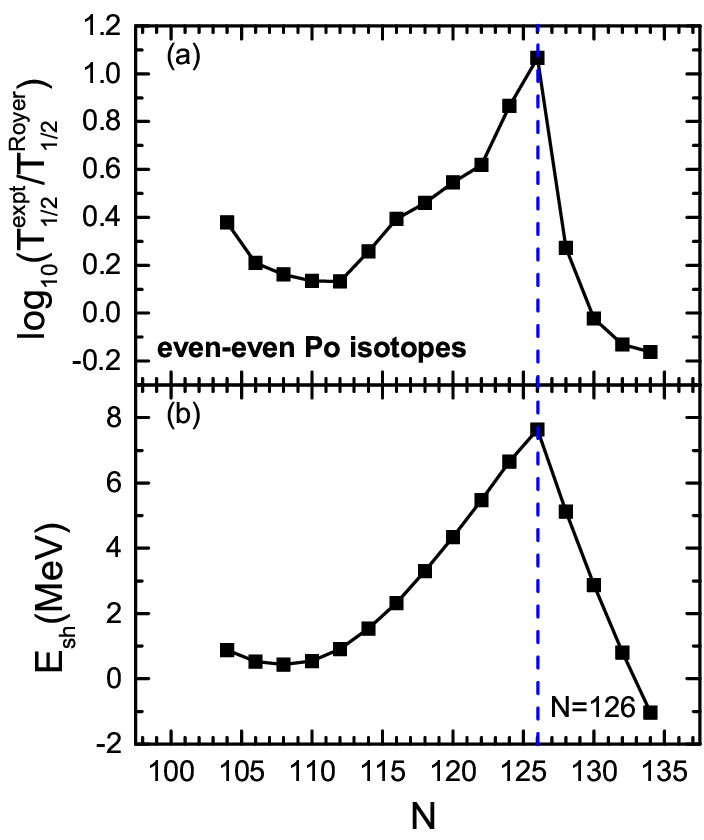}
	\caption{(a) The discrepancy between experimental and calculated logarithmic $\alpha$-decay half-lives using Eq.~(\ref{Eq1}) for even-even polonium isotopes is plotted against the neutron number $N$. (b) Similar to (a), except that the ordinate represents the variation of shell-correction energy \({E}_{\rm sh}\) with the neutron number, with calculated values taken from Eq. (\ref{Eq3}). The dashed line indicates the neutron number \(N = 126\). The variation trends and structural features of the two are highly similar.}\label{fig1}
\end{figure*}

We first applied the Royer law to compute the $\alpha$-decay half-lives of \textcolor{blue}{16} even-even polonium (Po) isotopes. Figure \ref{fig1}(a) plots the logarithmic differences between experimental data and calculations. As shown, a significant deviation in \(\log_{10}(T_{1/2}^{\rm expt}/T_{1/2}^{\rm Royer})\) emerges near the magic number \(N=126\) in the even-even Po isotopic chain. This indicates inadequate accounting of shell effects in this region by the original Royer law, particularly around the neutron magic number \(N=126\). To resolve such discrepancies, researchers have modified empirical formulas by introducing shell-effect-related terms. For example, Wang \textit{et al.} improved accuracy in Ref. \cite{Wangzy2015} by including a phenomenological shell-correction factor for nuclei near shell closures, but their fixed constant $S=0.5$ cannot fully capture the nuanced relationship between structural effects and $\alpha$-decay half-lives. Additionally, the Royer law's segmented approach lacks physical unification and neglects angular-momentum contributions.

\begin{table*}[htb]
	\centering
	\footnotesize
	\caption{Parameters of the Royer law \cite{2000Alpha}}
	\label{tab:params1}
	
	\begin{tabular}{lccc}
		\hline\hline
		Nuclear Type & $a$ (s) & $b$ (s) & $c$ (s.$\sqrt{\mathrm{MeV}}$) \\
		\hline
		Even-even        & -25.31 & -1.1629 & 1.5864 \\
		Even-$Z$/odd-$N$ & -26.65 & -1.0859 & 1.5848 \\
		Odd-$Z$/even-$N$ & -25.68 & -1.1423 & 1.5920 \\
		Odd-odd          & -29.48 & -1.1130 & 1.6971 \\
		\hline\hline
	\end{tabular}
\end{table*}

To address these limitations of the Royer law, we introduce an improved formulation that incorporates a shell-correction energy term, a pairing term, and an angular-momentum term into the original model. The resulting expression for the $\alpha$-decay half-life reads: 
\begin{align}\label{Eq2}
	\log_{10}T_{1/2} = a + b A^{1/6} \sqrt{Z} + c \frac{Z}{\sqrt{Q}} + d\left\{E_{\rm sh} - \left[ (-1)^Z + (-1)^N \right] 	+\frac{l(l+1)}{2}\right\},
\end{align}
where the coefficients $a$, $b$, $c$, and $d$ are determined from a fit to the 550 experimental data points and are listed in Table~\ref{tab:params2}.
Compared to the original Royer law, Eq.~(\ref{Eq2}) incorporates three physically motivated additions. The first additional term, \({E}_{\rm sh}\), represents the value of shell-correction energy, which captures the structural influence of the parent nucleus in the decay process. This form has previously been employed in calculations of spontaneous fission half-lives \cite{Baoxj15,Karpov12}. The second term, \([(-1)^Z + (-1)^N]\), acts as a unified pairing term that accounts for odd-even staggering across different nuclear parity combinations. It enables a consistent treatment of even–even, odd-$A$, and odd-odd nuclei within a single framework. Specifically, along an isotopic chain, even-even nuclei generally exhibit shorter half-lives than the average of their neighboring odd-$A$ nuclei, whereas odd-odd nuclei tend to have longer ones. This behavior is encapsulated in the formula as a correction of \(-2d\) for even-even nuclei, \(+2d\) for odd-odd nuclei, and $0$ for odd-$A$ nuclei. Such a unified approach not only reduces the number of free parameters but also offers a more consistent description of pairing effects. The third term, proportional to \(l(l+1)\), accounts for the orbital-angular-momentum carried by the emitted $\alpha$ particle, as commonly adopted in Royer-type formulas \cite{2020gyy,Wangzy2015,Ismail2022}. It is noteworthy that when an independent coefficient $f$ is introduced for this term, the fitting yields \(f \approx d/2\), supporting the current form. Collectively, these modifications yield a phenomenological yet physically grounded representation of the microscopic factors governing $\alpha$-decay systematics. The improved Royer formula thus provides a unified treatment of shell stabilization, pairing correlations, and angular-momentum hindrance within a single coherent framework.

\begin{table*}[htb]
	\centering
	\footnotesize
	\caption{Parameters of the improved Royer law}
	\label{tab:params2}
	
	\begin{tabular}{cccc}
		\hline\hline
		$a$ (s) & ~~~~$b$ (s) & ~~~~$c$ (s.$\sqrt{\mathrm{MeV}}$) & ~~~~$d$ (s) \\
		\hline
		$-28.1919\pm0.1510 $ &~~~~ $-1.0853\pm 0.0055$  &~~~~  $1.6260 \pm0.0039$  &~~~~  $0.1078 \pm 0.0026$  \\
		\hline\hline
	\end{tabular}
\end{table*}

\section{\label{sec:level3}THE RESULTS AND DISCUSSIONS}

From Fig. \ref{fig1}(a), we observe that the shell effects play a crucial role for certain nuclei, particularly near $N$ = 126. To incorporate shell effects into the Royer law, we first analyze the relationship between shell-correction energy ${E}_{\rm sh}$ and neutron number $N$ for even-even polonium isotopes in Fig. \ref{fig1}(b). The trend of ${E}_{\rm sh}$ closely mirrors the discrepancy between experimental and calculated $\alpha$-decay half-lives using the Royer law, with both peaking at $N = 126$. This suggests a linear correlation between ${E}_{\rm sh}$ and $\log_{10} T_{1/2}$, well described by the fit \(0.113E_{sh}+0.026\) for even-even Po isotopes. We therefore introduce a shell-correction energy \(d{E}_{\rm sh}\) into the original Royer law, where \({E}_{\rm sh}\) represents microscopic fluctuations of the nuclear binding energy relative to the macroscopic liquid-drop model \cite{Esh1}. It is computed as the difference between the experimental binding energy $B_{\text{expt}}$ and the macroscopic binding energy $B_{\text{LD}}$ of the nucleus,
\begin{eqnarray}\label{Eq3}
E_{\rm sh} =\frac{B_{\rm{expt}}-B_{\rm{LD}}}{1 \rm MeV}
\end{eqnarray}
Here the experimental binding energy is derived from the AME2020 \cite{AME2020} mass table, and $B_{\rm LD}$ is the theoretical binding energy of the spherical nucleus based on the liquid drop model, expressed as
\begin{align}\label{Eq4}
	B_{\rm LD} &= a_{v} A - a_{s} A^{2/3} - a_{a} \left( \frac{A}{2} - Z \right)^{2} / A  - a_{c} \frac{Z^{2}}{A^{1/3}} + a_{p} \delta A^{-1/2} ,
\end{align}
where  \(a_{v}\), \(a_{s}\), \(a_{a}\), \(a_{c}\) and \(a_{p}\) are the volume, surface, symmetry, Coulomb, and pairing energy coefficients, respectively. $\delta = +1$ for even-even nuclei, $\delta = 0$ for odd-$A$ nuclei, and $\delta = -1$ for odd-odd nuclei. The experimental binding energies of 2463 atomic nuclei ($Z \geq 8$, $N \geq 8$) selected from the AME2020 mass table \cite{AME2020} were fitted using the least squares method. The resulting parameters are: $a_{v} = 15.5287$ MeV, $a_{s} = 16.9043$ MeV, $a_{a} = 91.9686$ MeV, $a_{c} = 0.7025$ MeV, and $a_{p} = 12.4439$ MeV. The root-mean-square deviation of the fit is 3.02 MeV.

To evaluate the improvement brought by the inclusion of shell-correction energy and pairing effects, we compare the half-life predictions of the original Royer law (Eq. (\ref{Eq1})) with those of the improved version (Eq. (\ref{Eq2})). The accuracy of the calculations is quantified using the root-mean-square deviation (RMSD) between experimental and theoretical half-lives, defined as:
\begin{eqnarray}\label{Eq6}
	\sigma=\left[\frac{1}{m}\sum_{i=1}^{m} (\log_{10}T_{1/2}^{\mathrm{expt},i}-\log_{10}T_{1/2}^{\mathrm{cal},i})^2 \right]^{1/2},
\end{eqnarray}
where \(m\) denotes the number of nuclei considered in each case. A smaller RMSD corresponds to better agreement with experimental data and thus indicates improved model performance.

\subsection{\label{sec:level3}Favored $\alpha$ decay}
\begin{figure*}[htb]
	\centering
	\includegraphics[width=0.95\linewidth]{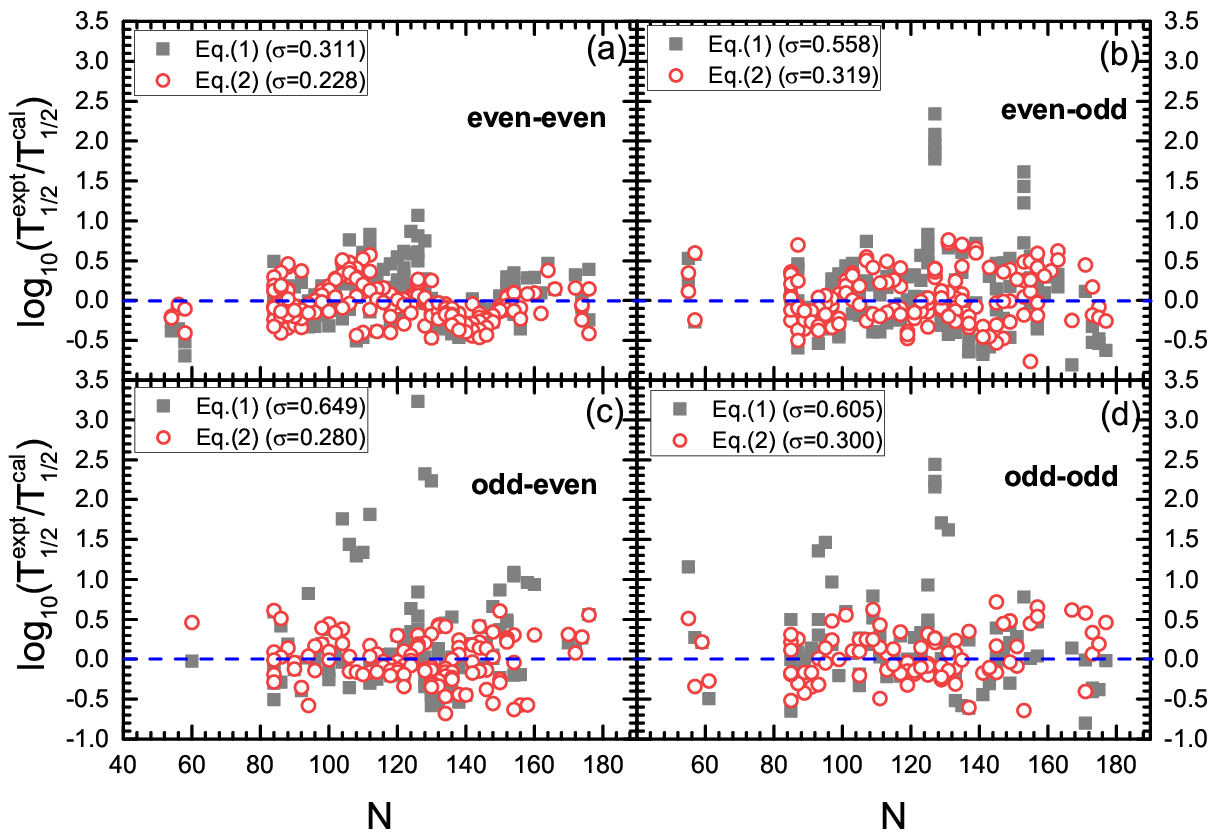}
	\caption{(Color online) Comparison of the differences between experimental and theoretical $\alpha$-decay half-lives calculated by Eq.~(\ref{Eq1}) and Eq.~(\ref{Eq2}) for four categories of $l=0$ nuclei: (a) even-even (190 nuclei), (b) even-odd (88 nuclei), (c) odd-even (71 nuclei), and (d) odd-odd (57 nuclei), plotted against neutron number $N$. The RMSD $\sigma$ is indicated in parentheses after each formula.}	\label{fig2}
\end{figure*}

Fig.~\ref{fig2} compares the deviations between experimental and calculated $\alpha$-decay half-lives for 406 favored $\alpha$ decays ($l= 0$) nuclei, using the original Royer law Eq.~(\ref{Eq1}) and  its improved revision Eq.(\ref{Eq2}). The dataset spans four decay types: (a) even-even (190 nuclei), (b) even-odd (88 nuclei), (c) odd-even (71 nuclei), and (d) odd-odd (57 nuclei))---a division necessitated by the Royer law's requirement for separate parameter sets per parity category.
In contrast, the pairing term in Eq.~(\ref{Eq2}) obviates the need for distinct parameter sets, allowing a single parameterization to uniformly describe all categories without loss of accuracy. Across all subsets, Eq.~(\ref{Eq2}) exhibits consistently smaller deviations than Eq.~(\ref{Eq1}), with the RMSD for $l=0$ nuclei reduced from 0.311 to 0.242 (Fig.~\ref{fig3}(a)). Most nuclei (open circles) show $\log_{10}(T_{1/2}^{\mathrm{expt}}/T_{1/2}^{\mathrm{cal}})$ values within [-0.4, 0.8] and cluster near the dotted line. Notably, the shell-correction energy term $E_{\text{sh}}$ captures partial nuclear structure effects, allowing Eq.~(\ref{Eq2}) to reproduce experimental values far more accurately near the neutron shell closure $N=126$, whereas Eq.~(\ref{Eq1}) exhibits pronounced discrepancies (solid squares).

\subsection{\label{sec:level3}Unfavored $\alpha$ decay}
For unfavored $\alpha$-decay ($l\neq 0$), the centrifugal potential barrier effect must be considered. This barrier, originating from the orbital-angular-momentum $l$ of the emitted $\alpha$ particle, reduces the tunneling probability and thus increases the half-life. Its contribution to the half-life can be directly incorporated into the Royer law via the angular-momentum term ${l(l+1)}$. The value of $l$ is determined by angular-momentum and parity conservation, as given by Eq.~(\ref{Eq5}). Notably, while selection rules permit multiple possible values for the orbital-angular-momentum $l$ of emitted $\alpha$-particles, we adopt the minimum allowable value $l_{\text{min}}$ in all subsequent calculations for simplicity.
\begin{eqnarray}\label{Eq5}
	l_{min}=\left\{
	\begin{array}{llr}
		\Delta _{j}, ~~~~~~~for~~even~~\Delta _{j}~~and ~~\pi _{p}=\pi _{d}\\
		\Delta _{j}+1, ~~for~~even~~\Delta _{j}~~and ~~\pi _{p}\ne \pi _{d}\\
		\Delta _{j}, ~~~~~~~for~~odd~~~~\Delta _{j}~~and ~~\pi _{p}\ne\pi _{d}\\
		\Delta _{j}+1, ~~for~~odd~~~~\Delta _{j}~~and ~~\pi _{p}= \pi _{d}\\
	\end{array}
	\right.
\end{eqnarray} 
where $\Delta_{j}=\mid j_{p}-j_{d}\mid$, with $j_{p}$, $\pi _{p} $, $j_{d}$ and $\pi _{d} $ represent the spin and parity values of the parent and daughter nuclei, respectively. Their values used in this work are taken from Refs. \cite{Nubase2020,Goriely13}.

\begin{figure*}[htb]
	\centering
	\includegraphics[width=1.0\textwidth]{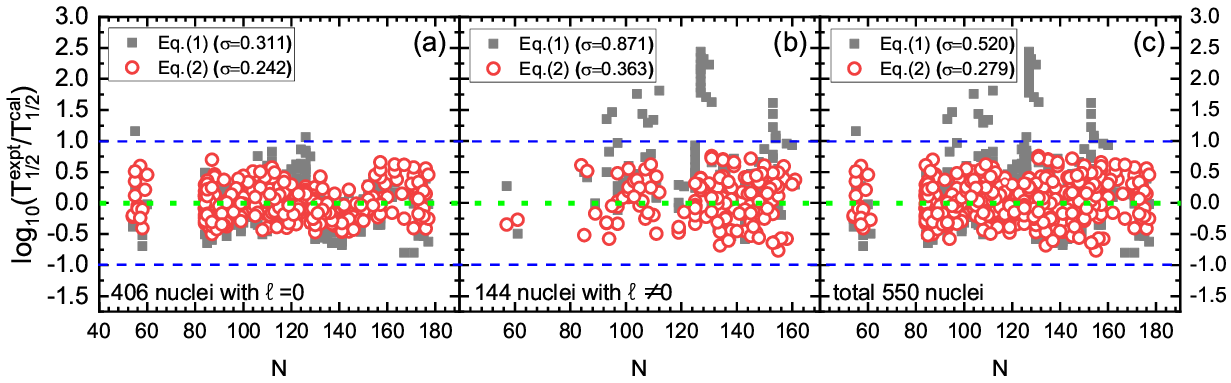} 
	\caption{(Color online) Comparison of deviations between experimental and calculated $\alpha$-decay half-lives: (a) 406 favored $\alpha$-decays with angular-momentum $l = 0$; (b) 144 unfavored $\alpha$-decays with $l \neq 0$; (c) Full dataset of 550 nuclei.}	\label{fig3}
\end{figure*}

To evaluate the effect of the third additional term, we compare calculations with and without the angular-momentum $l(l+1)$ term in Eq. (\ref{Eq2}) for 144 unfavored $\alpha$-decay half-lives ($l\neq 0$). The resulting RMSDs are 0.363 (with $l(l+1)$) and 0.871 (without it), indicating the importance of explicitly including the $l(l+1)$ contribution. Including this term reduces the RMSD from 0.871 (Eq.~(\ref{Eq1})) to 0.363, corresponding to a 58.3\% improvement in accuracy. In these cases, a nonzero  $l$ introduces a centrifugal barrier in unfavored  $\alpha$ decays---an effect that is not accounted for in the absence of the $l(l+1)$ term. Incorporating this term effectively captures the centrifugal barrier effect. This correction is especially significant for unfavored $\alpha$ decays ($l\neq 0$) in odd-$A$ and odd-odd nuclei, leading to improved accuracy in half-life predictions across all nuclear types. The good agreement between the calculated results (Fig.~\ref{fig3}(c)) and experimental data demonstrates that the improved Royer law, which includes both the shell-correction energy and the angular-momentum $l(l+1)$ term, performs well across all 550 $\alpha$-decay cases studied.

\begin{table*}[htb]
	\centering
	\footnotesize
	\caption{RMSDs of five empirical formulas.}
	\label{tab:params4}
	\vskip 2mm
	\tabcolsep 4.5pt
	\begin{tabular}{cccc}
		\hline\hline
		& \( l_{\text{min}} = 0 \) & \( l_{\text{min}} \neq 0 \) & Total \\
		& (n=406) & (n=144) & (n=550) \\
		\hline
		Eq.~(\ref{Eq1}) & 0.311 & 0.871 & 0.520 \\
		Eq.~(\ref{Eq2}) & 0.242 & 0.363 & 0.279 \\
		Ref.~\cite{2020gyy} & 0.322 & 0.443 & 0.358 \\
		Ref.~\cite{Wangzy2015} & 0.314 & 0.385 & 0.334 \\
		Ref.~\cite{Ismail2022} & 0.300 & 0.581 & 0.393\\
		\hline
	\end{tabular}
\end{table*}
The calculations of $\alpha$-decay half-lives with three other well-known empirical formulas in Refs.~\cite{2020gyy,Wangzy2015,Ismail2022} are also performed and the corresponding RMSDs are presented in Table ~\ref{tab:params4}.
we evaluate separately for nuclei with \( l = 0 \), \( l\ne 0 \), and the full dataset of 550 nuclei. 
Comparing the results, it is found that the improved Royer law Eq.~(\ref{Eq2}) yields the smallest values of RMSDs for both the full data set (0.279) and for two subsets (0.242 for favored and 0.363 for unfavored). In other words, the precision in our formula is better than that of the previous methods.

\begin{figure}[htbp]
	\centering
	\includegraphics[width=0.9\textwidth]{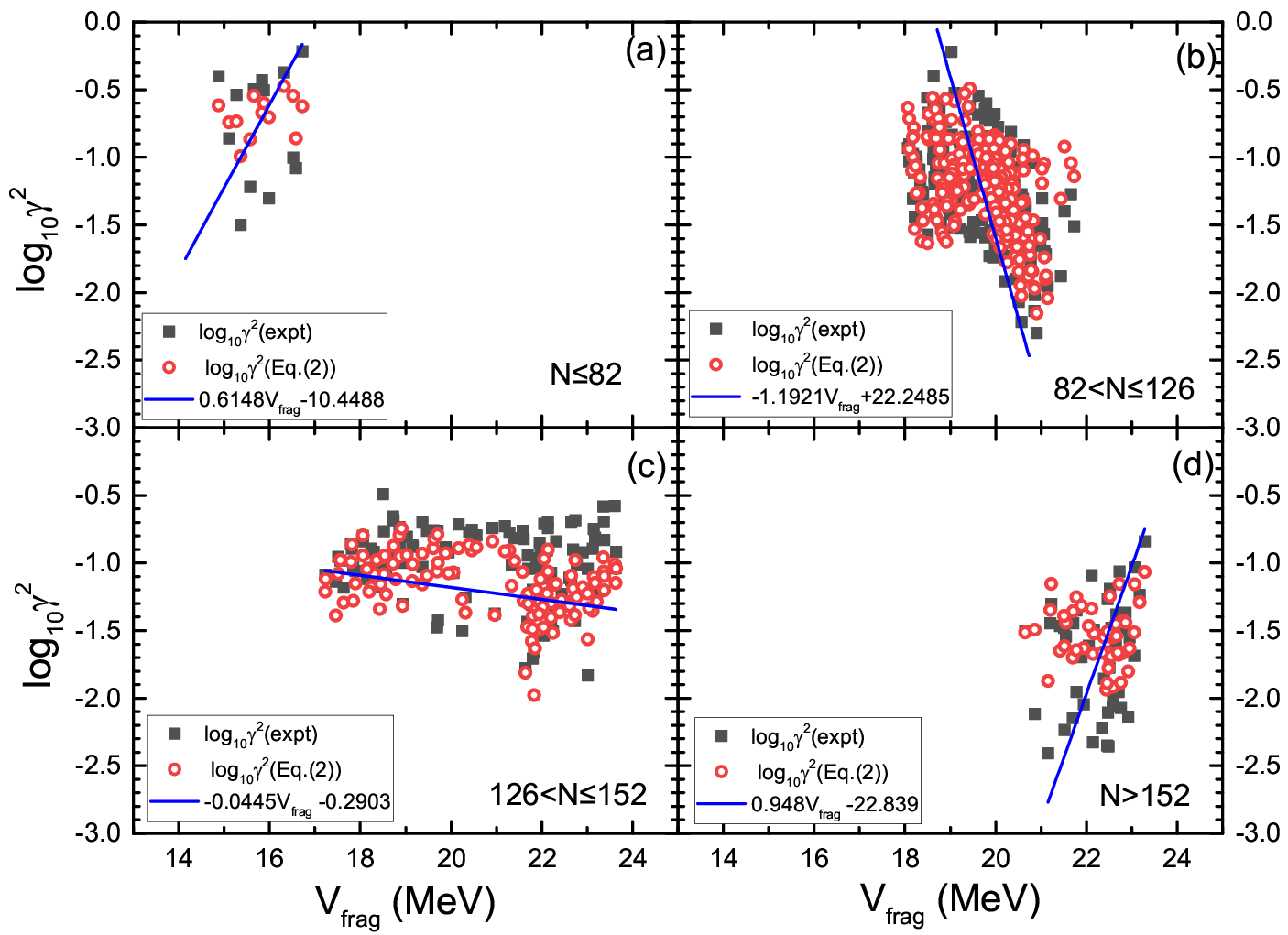}
	\caption{Systematic behavior of $\log_{10}\gamma^2$ as a function of the fragmentation potential $V_c - Q$. Panels (a) to (d) correspond to the neutron-number regions: (a) $N \leq 82$, (b) $82 < N \leq 126$, (c) $126 < N \leq 152$, and (d) $N > 152$. Black squares represent values derived from experimental half-lives, and red circles denote results calculated using the improved Royer law in Eq.~(\ref{Eq2}). The blue lines indicate the linear fitting of the experimental data.}
	\label{fig4}
\end{figure}

To further examine the physical reliability of the improved Royer law, we performed a systematic analysis of the reduced $\alpha$-decay widths. The reduced width $\gamma^2$ is defined as \cite{VQ}
	\begin{equation}
		\gamma^2 = \frac{\Gamma}{2P} \, ,
	\end{equation}
	where $\Gamma$ is the decay width and $P$ is the Coulomb penetrability evaluated.
	The decay width is related to the half-life $T_{1/2}$ by
	\begin{equation}
		\Gamma = \frac{\hbar \ln 2}{T_{1/2}} \, .
	\end{equation}
	Hence, the logarithm of the reduced width can be expressed as
	\begin{equation}
		\log_{10}\gamma^2 = \log_{10}(\hbar \ln 2) - \log_{10}T_{1/2} - \log_{10}P-\log_{10}2.
	\end{equation}
	The Coulomb barrier at the touching configuration is given by
	\begin{equation}
		V_c(r_B) = \frac{Z_d Z_\alpha e^2}{r_B}, \quad 
		r_B = 1.2\,(A_d^{1/3} + A_\alpha^{1/3})~\text{fm},
	\end{equation}
	where $Z_d$ and $A_d$ denote the proton and mass numbers of the daughter nucleus, and $Z_\alpha=2$, $A_\alpha=4$ for the $\alpha$ particle. 
	The fragmentation potential is then defined as
	\begin{equation}
		V_{\text{frag}} = V_c(r_B) - Q.
\end{equation}
The analysis focuses on 406 favored $\alpha$-decay nuclei with $l=0$, where the centrifugal barrier is absent, thereby providing a clear test of the linear relationship between $\log_{10}\gamma^2$ and $V_{\mathrm{frag}}$. Fig.~\ref{fig4} presents the results across four neutron-number regions, with experimental data indicated by black squares and theoretical values from Eq.~(\ref{Eq2}) shown as red circles. As observed, both experimental and theoretical results adhere to the expected nearly linear trend between $\log_{10}\gamma^2$ and $V_{\text{frag}}$ in each region. This agreement confirms that the improved formula not only enables more accurate predictions of half-lives but also establishes an approximate linear relationship between reduced widths and fragmentation potentials. The robustness of this correspondence further validates the physical reliability of our model across different nuclear regions. Although minor local fluctuations are present, the overall linearity between $\log_{10}\gamma^2$ and $V_{\mathrm{frag}}$ remains clear, in line with the universal behavior reported by Delion \cite{VQ}.

\begin{figure*}[htb]
	\centering
	\includegraphics[width=1.0\textwidth]{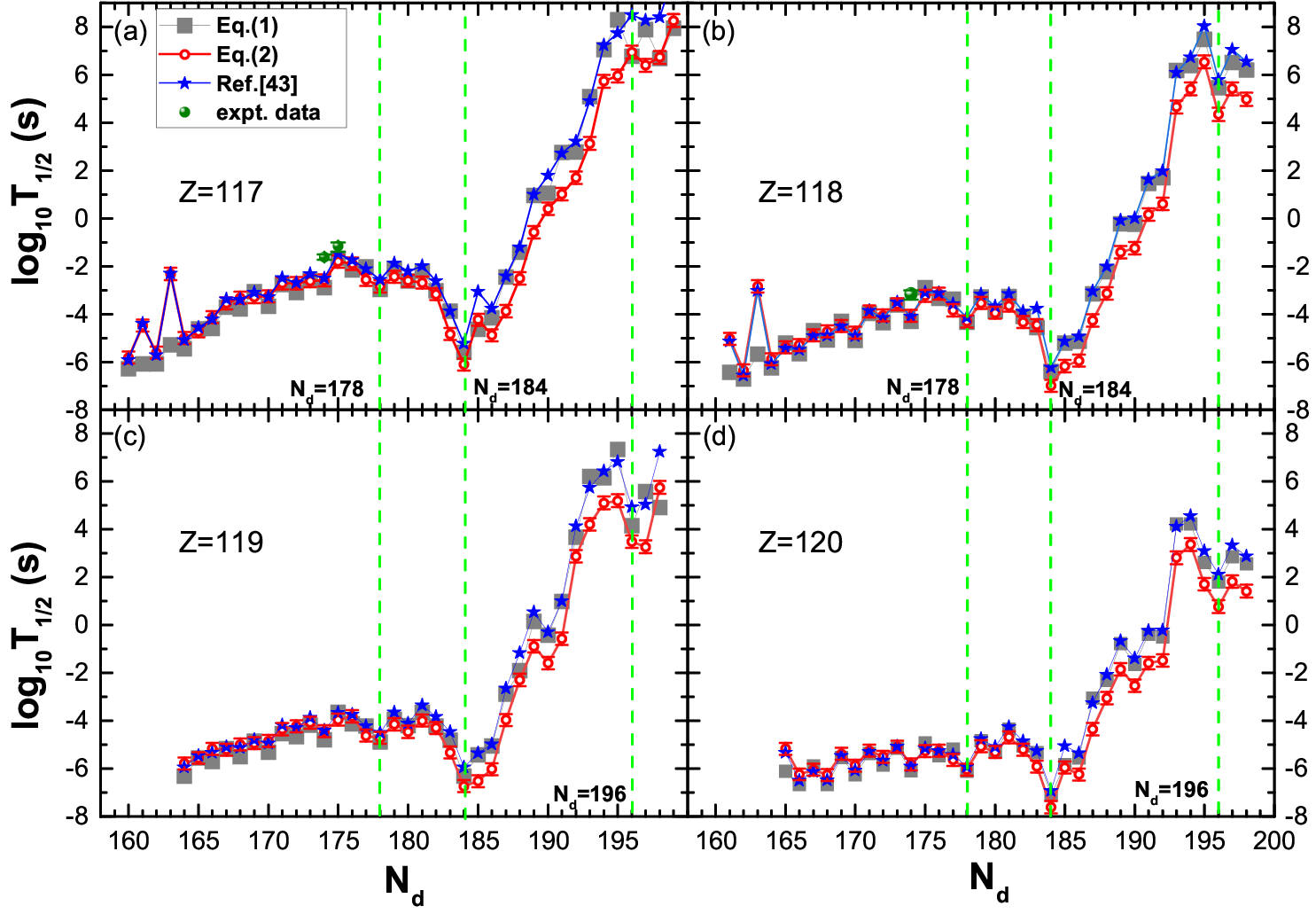}
	\caption{(Color online) The $\log_{10} T_{1/2}$ values of $Z=117-120$ isotopes versus neutron number of daughter nucleus $N_d$. The open circles, stars, and the solid squares denote the prediction results with the improved Royer law (Eq.~(\ref{Eq2})), the original revision (Eq.~(\ref{Eq1})), and the DZR model (Ref.~\cite{2020gyy}). Experimental data (solid circles) for $^{293,294}\mathrm{Ts}$ and $^{294}\mathrm{Og}$ are included for comparison. The vertical dashed lines indicate the possible existence of magic numbers or neutron subshell structures at neutron numbers $N_d = 178$, 184, and 196.}	\label{fig5}
\end{figure*}

To further verify the applicability of the improved formula (Eq.~(\ref{Eq2})), we used it to calculate the $\alpha$-decay half-lives of superheavy nuclei with $Z = 117$–$120$. For those nuclei lacking experimental $Q$-values or binding energies, we adopted the Weizsäcker–Skyrme (WS4+RBF) mass table \cite{WS4}. In such cases, the shell-correction energy $E_{\mathrm{sh}}$ is evaluated using binding energies from the WS4+RBF model, where the experimental binding energy $B_{\mathrm{expt}}$ is replaced by $B_{\mathrm{WS4+RBF}}$ to maintain consistency with the definition given in Eq.~(\ref{Eq3}). Fig.~\ref{fig5} presents the calculated $\alpha$-decay half-lives as a function of the daughter neutron number $N_d$, using the improved Royer law (Eq.~(\ref{Eq2})), the original Royer law (Eq.~(\ref{Eq1})), and the DZR model \cite{2020gyy}. Although three different computational methods were employed, they all exhibit the same variation trend and consistently indicate the possible existence of magic numbers or neutron subshell structures at neutron numbers $N_d = 178$, 184, and 196. As shown, the improved Royer law yields results that are more consistent with the systematic trends predicted by the DZR model in the region $N_d \le 184$, and lower than those results of two in the region $N_d \ge 184$. 
This consistency is particularly evident for the existing experimental data $^{293,294}$Ts and $^{294}$Og. Overall, the improved formula demonstrates strong extrapolation capability in predicting the decay properties of superheavy nuclei, reinforcing its reliability beyond the region used for parameter fitting.

\begin{table*}[htb]
	\centering
	\footnotesize
	\caption{Comparison of model performance: Royer law versus its improved revision.}
	\label{tab:params5}
	
	\begin{tabular}{lcc}
		\hline\hline
		Metric & Eq.~(\ref{Eq1})  & Eq.~(\ref{Eq2}) (Improved) \\
		\hline
		Number of parameters    & 12               &4               \\
		RMS deviation           & 0.520            & 0.279         \\
		Parity treatment        & Segmented        & Unified          \\
		Physics extensions      & None             & Shell+pair+$l$ term \\
		\hline
	\end{tabular}
\end{table*}

\section{\label{sec:level4}Summary}
In summary, we have developed an improved Royer formula, Eq.~(\ref{Eq2}), for calculating $\alpha$-decay half-lives by incorporating three physically motivated correction terms—shell-correction energy, pairing effects, and angular momentum—into a unified four-parameter framework. This work not only offers a simplified and more accurate empirical formula but also establishes a structure that naturally integrates nuclear-structure corrections, thereby bridging phenomenological approaches with microscopic insights. Unlike the original Royer law, which treats nuclei differently based on parity, the new formulation provides a unified description for all nuclei, reducing the number of free parameters from 12 to 4—a 66.7\% reduction in complexity. The inclusion of shell-correction energy markedly mitigates discrepancies near the neutron number $N = 126$ and improves the overall predictive accuracy of $\alpha$-decay half-lives. Moreover, the angular-momentum term accounts for hindrance effects arising from spin and parity changes between parent and daughter nuclei. As a result, the model consistently describes both favored and unfavored $\alpha$-decays within a single framework, improving physical coherence and practical utility. Using this refined formula, we systematically computed the half-lives of 550 $\alpha$ transitions between ground states of parent and daughter nuclei, achieving a significant improvement in accuracy: the root-mean-square deviation drops from 0.520 to 0.279, corresponding to a 46.3\% enhancement (see Table~\ref{tab:params5}). We further applied the formula to predict $\alpha$-decay half-lives for superheavy nuclei with $Z = 117$–$120$. The formula also captures the emergence of magic numbers or neutron subshell structures at neutron numbers $N_d = 178$, 184, and 196.

\begin{center}
\textbf{ACKNOWLEDGMENTS}
\end{center}

This work was supported by the Guangxi Natural Science Foundation (Nos. 2023GXNSFDA026005, and 2023GXNSFBA026008), the National Natural Science Foundation of China (Nos. 12465019, 12465021, 12265006 and U1867212), and the Central Government Guides Local Scientific and Technological Development Fund Projects (No. Guike ZY22096024).

\end{document}